# Swelling, Softening and Elastocapillary Adhesion of Cooked Pasta

Jonghyun Hwang[1*], Jonghyun Ha[1*], Ryan Siu[1], Yun Seong Kim[1,2], and Sameh Tawfick[1,2†]

[1]Department of Mechanical Science and Engineering, University of Illinois at Urbana-Champaign, Urbana, Illinois, 61801, USA
[2]The Beckman Institute for Advanced Science and Technology, University of Illinois at Urbana-Champaign, Urbana, Illinois, 61801, USA

*These authors contributed equally to the work
†Author to whom correspondence should be addressed. E-mail: tawfick@illinois.edu

**Abstract**

The diverse chemical and physical reactions encountered during cooking connect us to science every day. Here, we theoretically and experimentally investigate the swelling and softening of pasta due to liquid imbibition, as well as the elastic deformation and adhesion of pasta due to capillary force. As water diffuses into the pasta during cooking, it softens gradually from the outside inward as starch swells. The softening follows three sequential regimes: Regime I shows a slow decrease of modulus with cooking time; Regime II, the glassy to rubbery transition region, is characterized by very fast, several orders of magnitude drop in modulus; and regime III, the rubbery region, has an asymptotic modulus about four orders of magnitude lower than the raw pasta. We present experiments and theory to capture these regimes and relate them to the heterogeneous microstructure changes associated with swelling. Interestingly, we observe a modulus drop of two orders of magnitude within the range of "al dente" cooking duration, and we find the modulus to be extremely sensitive to the amount of salt added to the boiling water. While most chefs can gauge the pasta by tasting its texture, our proposed experiment, which only requires a measurement with a ruler, can precisely provide an optimal cooking time finely tuned for various kinds of pasta shapes.

**Main text**

Pasta cooking offers intriguing chemo-mechanical and hydrostatic and -dynamic phenomena for the curious mind. Examples of such phenomena manifest themselves due to interaction between soft elastic solids and liquid such as in steaming dumplings, frying potatoes, and swelling of noodles. Liquid and heat transport into cooked media can change the materials' shape and mechanical properties.[1–4] Another common example is seen with starch, the most abundant carbohydrate in human diets, which expands and softens as it experiences hygroscopic swelling.[5–8] In addition to swelling, capillary attraction and adhesion is commonly encountered during cooking of slender forms like noodles, as well as during serving and eating of noodle soups. Complex mechanics laws govern the interactions between fully cooked noodles hanging from a fork and determine how they stick together or fall back down (hopefully) into the bowl.

In this article, we document how pasta, composed of starch granules,[9,10] swells and softens as a function of the cooking duration, water temperature, and the addition of salt. As expected, noodles not only experience volumetric expansion and but also their bending rigidity (elastic modulus multiplied by second moment of area) significantly decrease so that noodles become edible. We focus on a simple but insightful kitchen experiment wherein softened pasta noodles coalesce when the noodle strings are lifted, as demonstrated in Fig. 1(a). The coalescence depicted in the figure is mainly due to increased flexibility. Fig. 1(b) shows the time evaluation of hygroscopic swelling of the pasta noodle. Notably, there is a significant increase in length as well. The diameter to length relative swelling anisotropy is 3.5. We break the cooked noodles at various cooking time to observe the cross section. As it's cooked, the raw pasta soon become composed of hard inner regions and softened, cooked outer layers (traditionally expected structure of Al Dente). This soft to hard transition in the radial direction of pasta can be explained by water transport and diffusion starting from the outermost surface of the noodle, hence the softening initiates from outermost region. It then reaches the saturation stages, where we observe no color gradient in the cross section (fully soft, and arguably overcooked). To study the elasticity of cooked pasta, we solve the liquid diffusion equation to obtain an expression for the concentration of liquid within the pasta and use it to develop the theoretical model of the swelling dynamics of starch materials. The concentration profile is also used to model the elastic modulus $E$ of the material based on the pioneering work of Weibull.[11] Combining the two perspectives, swelling and softening, we construct a predictive model for pasta elastocapillarity, namely the stick length of the noodles due to capillary forces.[12–18] Overall, we believe that not only is this one of the first studies combining swelling dynamics with elastocapillarity, but also this kitchen experiment is insightful for understanding pasta cooking and for teaching mechanics.

We physically explain the capillary-driven deformation of slender structures, termed elastocapillarity[15], before turning our attention to imbibition and swelling. When slender structures such as thin rods or plates are pulled out of a liquid container, the liquid surface energy causes coalescence. Fig. 2 shows the elastocapillary adhesion of two noodles seen during the experiment, 2(a), and the schematic representation of the phenomenon, 2(b). A liquid meniscus forms between the coalesced structure (see inset (i) of Fig. 2(b)), and this capillary contact balances the elastic resistance created by the bending of the elastic rods. The adhesion energy created by the surface tension of liquid can be expressed as $E_c \sim \gamma p(l_0 - l_s)$, where $\gamma$, $p$, $l_0$, and $l_s$ are the surface tension, perimeter of the wetted surface, length of the noodle, and the stick length (see inset (ii) of Fig. 2(b)), respectively. If the rods are displaced by a distance $d$, as shown in Fig. 2(a), the elastic resistance energy to the surface tension can be expressed as $E_e \sim EId^2/l_s^3$, where $I$ is the second moment of area. The stick length is defined as the distance between the fixture and point where the meniscus between the pasta forms.[14,15] Then, an expression for the stick length can be obtained by balancing the two forces, $F_c$ and $F_e$: $l_s \sim (dl_{ec})^{1/2}$, where $l_{ec} = [EI/(\gamma p)]^{1/2}$ is the elastocapillary length.[13,19] Since the Young's modulus, $E$, and other geometrical parameters ($I$ and $p$) vary over time and the surface tension $\gamma$ is a function of temperature[20], $l_{ec}$ is a function of temperature, cooking time and added salt as described below.

We now move on to finding an analytic expression for the time-variant geometry of the pasta. Volumetric expansion of the pasta is due to hygroscopic swelling, where migration of the water molecule into the pasta induces expansion of the starch matrix, as represented in Fig. 3(a). To capture how the geometrical factors change over time and their effect to the elastocapillary length, we define a length scale of the cross-section of pasta, $\delta$. Then, hygroscopically swelled $\delta$ can be written as $\delta \sim \delta_0(1+\epsilon)$, where $\delta_0$ is an initial length scale, and $\epsilon$ is the strain. In the hygroscopically swelling material, we assume that the expansion in volume is entirely due to the liquid migration into the material.[21,22] Hence, we solve the diffusion equation to relate the liquid concentration to the volumetric strain. The diffusion equation, $D\nabla^2 C = \partial C/\partial t$, is subject to the following boundary and initial conditions: $C(r,t)|_{r=R} = C_0$, $\partial C/\partial r|_{r=0} = 0$ and $C(r,t)|_{t=0} = 0$. Finally, we get a general expression for the liquid concentration within the solid.

$$C(r,t) = [c_0 J_0(\sqrt{\lambda_n} r)][c_1 e^{-D\lambda_n t}], \tag{1}$$

where $D$ is the diffusivity of the material, $\lambda_n$ is the $n^{th}$ eigenvalue of the system, $c_0$ and $c_1$ are coefficients determined by the initial and boundary condition of the system. We note that Eqn. (1) is composed of cylindrical Bessel function, $J_0$. The Bessel function is an oscillating function that decays over time. Because the system contains an infinite number of eigenvalues, the complete expression contains an infinite sum of the functions. (see Note S1 for the complete analytical expression for the liquid concentration). By integrating the liquid concentration profile, $C(r,t)$, within the pasta to get the amount of liquid migrated,[23] we obtain an expression of hygroscopic strain in radial direction, $\epsilon_{rr}(t)$, which is expressed in Eqn. (2). (See Note S1 for the detailed derivation).

$$\epsilon_{rr}(t) = \alpha \left(1 - 4\sum_{n=1}^{\infty} \frac{e^{-x_n^2 t/\tau}}{x_n^2}\right), \tag{2}$$

where $\alpha$ is the coefficient of expansion, $x_n$ is $n^{th}$ root of the cylindrical Bessel function of the first kind, and $\tau$ is the time constant $\tau = \beta R^2/D$. $R$ and $\beta$ are the initial radius of the past and the saturation time constant, respectively. Next, we replace the length scale, $\delta$, with the radius of the pasta noodle as a function of time, $r(t) = R[1 + \epsilon_{rr}(t)]$. Considering swelling effects of pasta, we conclude the geometrical factors in the elastocapillary length to be $I \sim (1+\epsilon)^4$ and $p \sim (1+\epsilon)$, so $I/p \sim (1+\epsilon)^3$.

The hygroscopic strain and elastic modulus of the pasta noodle at various cooking times and temperature were measured by cooking 2 mm-diameter spaghettis (brand name De Cecco). In our experimental window – maximum of 30 minutes of cooking, spaghetti noodles display radial strain and axial strain approximately of 70% and 20%, respectively. This difference is due to the difference in exposed surface area of pasta to the liquid where stiffer inner region of the pasta interrupt elongation in the axial direction. Pasta radius at saturation was 96% of the initial, uncooked radius, for both the cooking temperatures, 80 °C and 100 °C. We observed that presence of the salt ions in the water (Al dente recipe) facilitate hygroscopic swelling. Next $\alpha$ and $\beta$ are determined experimentally by relating that the maximum strain of the noodles, which is approximately 96%, to find that $\alpha = 0.9588$ and $\beta = 0.625$. The saturation time to reach the maximum radial strain differed by cooking temperature. For 100 °C cooking, radial growth saturates near $\tau \approx 105$ min and for 80 °C cooking, $\tau \approx 185$ min. The diffusivity of pasta was then found from this relation: $D = \beta R^2/\tau$. $D$ of 80 °C and 100 °C are $D = 5.682 \times 10^{-11}$ m$^2$/s and $D = 9.921 \times 10^{-11}$ m$^2$/s, respectively. Using each $D$, we compare our theoretical modeling with the experimental measurements of radial expansion at various cooking times, as shown in Fig. 3(b).

Liquid migration into the pasta not only swells it, but it also causes significant softening. To obtain the quantitative trend of pasta softening, we measure the elastic modulus of the pasta in various cooking times by using the Dynamic Mechanical Analyzer (DMA-840, TA instrument). We submerge the samples (noodles) in hot water and cook them for set cooking times, as shown in Fig. 3(c). The pasta is considerably rigid at an initial stage, but elastic modulus exponentially decreases as cooking time increases. Similar to what we observe in the modulus-temperature curve of polymers, we observe three regimes when we plot log($E$) versus time. In the Regime I, equivalent to the hard-glassy region, the modulus drops very slowly. In Regime II, where the glassy to rubbery transition occurs, the modulus quickly drops; three orders of magnitude within about four minutes. In Regime III, the rubbery region, the modulus asymptotically reaches a final value for the fully cooked pasta. Unexpectedly, the presence of salt ions in the water stiffens the material compared to the modulus of the noodle cooked in the same condition but without salt. Overall, the modulus drops by 5 orders of magnitude from the initial ($E_0 = 2.17$ GPa $\pm$ 0.15) to the saturated ($E_s \sim 10^2$ kPa). (see Fig. 3(d)) Similar initial values of elastic modulus of pasta noodle were reported in previous work.[24]

To mathematically describe Young's modulus variation with the cooking times (softening dynamics), we used the Weibull cumulative distribution of the form (CDF), $F(\bar{C} \mid \eta, \theta) = 1 - \exp[-(\bar{C}/\eta)^\theta]$, where $\bar{C}$ is the nondimensional liquid concentration, and $\eta$ and $\theta$ are scale and shape parameters of the Weibull CDF., respectively. (see Note S2 for the description of these parameters). Weibull statistics have been traditionally used in mechanics to represent fracture problems where it describes how bond ruptures in the secondary bonds inside the polymer network lead to orders of magnitude drop in the modulus during the glassy to rubbery state transition.[25,26] Previously, this state in transition, so called 'reptation,' as de Gennes puts it,[27] would sometimes be modeled using Boltzmann distribution, where it only took in the bond rupture activation energy to estimate the modulus. However, we agree with the Mahieux and Reifsnider's argument that the secondary bond rupture happen at different times "due to the distance variations between atoms inducing a distribution in the strength of the interactions," and hence adopt the Weibull statistics to model the modulus of the starch granules, which is a semi-crystalline polymer.[28] Using Weibull statistics for modeling modulus drop is hence also based on the relation between extreme swelling and bond rupture, but the details of these changes in molecular microstructure are outside the scope of this study.

Adopting Slade and Levine's temperature and moisture content equivalency,[29] the Weibull modulus can be written as a function of liquid concentration within the pasta.[30]

$$E_i(r,t) = \exp\left\{\ln E_0 - [\ln E_0 - \ln E_s]\left[1 - \exp\left(-\left(\frac{C(r,t)/C_0}{\eta}\right)^\theta\right)\right]\right\}. \quad (3)$$

where $E_0$ and $E_s$ correspond to the elastic modulus of the initial and saturation, respectively. $E_0$ and $E_s$ constrain the upper and lower boundary of the modulus in our experimental window, and are experimentally obtained: $E_0$ = 2.17 GPa, $E_s$ of 80 °C = 135 kPa, and $E_s$ of 100 °C = 50.8 kPa. The local modulus in Eqn. (3) is expressed as a function of local water concentration, $C(r,t)$ of Eqn. (1). This model allows us to analytically express the modulus of the pasta at any radial position at varying hydration, thus plasticization, rate. Our model advances the previous models for softening dynamics in that ours can be generally used when finding the modulus of polymers that uses liquid as plasticizer at specific time and location within the polymer matrix.

We can find $\eta$ and $\theta$ by comparing the area-averaged modulus to the experimentally measured modulus, By integrating Eqn. (3) over the area, we obtain the area-averaged Young's modulus:

$$E(t) = \frac{1}{\pi R^2}\int_0^{2\pi}\int_0^R E_i(r,t)r\,dr\,d\theta. \quad (4)$$

Then, by fitting the centroid of the local modulus to the measured elastic modulus, one can back-solve for the parameters $\eta$ and $\theta$. We report the $\eta$ to be 0.077 and $\theta$ to be 0.83, same for both the cooking temperatures. The experimentally measured young's modulus of the composite and the analytic expression of the modulus are drawn in Fig. 3(d).

Substituting the expressions found in Eqns. (2) and (3) to the expression of $l_{ec}$ gives the following expression of the stick length as function of the measured parameters

$$l_s \sim \left\{d\left[\frac{\sum_{i=0}^{\infty} E_i(r,t)[r_{i+1}(t)^4 - r_i(t)^4]}{8\gamma(T)\cdot r(t)}\right]^{1/2}\right\}^{1/2}. \quad (5)$$

Here, the subscript $i$ denotes the $i^{th}$ co-axial cylinder whose inner diameter equals $r_i(t)$, outer diameter equals $r_{i+1}(t)$, and the modulus equals $E_i(r,t)$. We assume the liquid wets each half of the cylinder perimeter[19], hence the wicking perimeter $p$ is $2\pi r(t)$. Furthermore, the surface tension of our cooking liquid, water, is assumed constant throughout the experiment; although the starch concentration in the water bath would gradually increase, because the change in the concentration of starch in the water bath is very small ($dC_{solute} \sim 0$), we assume a constant surface tension for each cooking temperature. For 80 °C, the surface tension of water is 62.67 $mN/m$ and for 100 °C, it is 58.91 $mN/m$.[31]

To validate our theoretical model, Eqn. (5), we experimentally measured $l_s$ in various cooking times, as shown in Fig. 4(a). The stick length experiments were done by submerging and then subsequently removing pasta noodle from a heated water bath conditioned at either 80 °C or 100 °C. Pasta noodles mounted at the top were cooked up to 30 minutes, and noodles were drained out of the bath at set cooking times (every 3 minutes) to measure the stick length. We plot the experimental data and the analytic solution for the stick length in Fig. 4(b). The prefactor for the scaling law, $\psi$, is used to precisely fit the experimental results: $l_s = \psi(dl_{ec})^{1/2}$, where $\psi = 1.25$.

We observed that $l_s$ initially drops exponentially until it reaches equilibrium. The capillary-induced adhesion starts during Regime II of $\log(E)$ versus time. Interestingly, even with an increment of the radius of pasta, $l_s$ exponentially decreases at early cooking times. This is because the rigidity, $B$, is determined by not only the size of structure but also the elastic modulus of it; in the case of pasta, the decrease in modulus is much larger than the increase of strain triple powered ($I/p$) at early cooking times. This trend in $l_s$ agrees with our expectation; because the modulus drops quickly initially, $l_s$ decreases quickly at first, and it levels down eventually as cooking time increases. Hence, $l_s$ is highly modulus dependent. However, one can expect the stick length to slightly increase with an extended cooking time. We observed that the modulus saturates before the strain saturates. Once the pasta is fully saturated with the liquid and there is negligible gradient in the liquid concentration within the liquid pasta, the pasta can then be treated as a single material. As the noodle's radius will still increase even after full penetration of water to the center of pasta, the rigidity, $B$, will increase, making $l_s$ to increase over time until $\tau$. This aspect is seen at both the analytic curves in Fig 4(b). In the model, the minima occur near 26 minutes and 15 minutes of cooking for 80 °C and 100 °C, respectively. This expectation is not reflected on the experimental data because the increased volume, hence weight, of the pasta acts against the elastic resistance of deformed pasta. However, the general trend of the capillary adhesion is well captured even without the gravitational effect, for which we did not consider for the simplicity of the model.

Adding salt in the cooking water, which many cooking books advise for better pasta texture, affects the chemical and mechanical properties of the pasta in some interesting ways. We observed increase in both the strain (swelling) and modulus (stiffening) of the noodles cooked in salted water compared to those cooked in distilled water, as shown in the star symbols of Fig. 3(b) and (d). Increase in the rate of hygroscopic swelling can be related to the facilitated transport of the "hydrated ions" into the polymer. Some studies reported increase in water/ion mobility into polymer matrix with increase in the salt concentration.[32] This would mean there would be less modulus gradient within the solid as water diffusion is faster with ions. Increase in the modulus is attributable to the increased van der Waals attraction induced by the presence of salt ions between the macromolecular chains. Thus, plasticization by hydration occurs slower. Our interpretation is that addition of salt would provide more homogenous, and unique texture due to ionic interaction.

We were surprised to realize that within the few minutes cooking range of the al dente recipe recommended by various manufacturers, the modulus drops by two orders of magnitude. This necessitates great experience in cooks to taste the right texture of al dente, and as expected causes great variation. We also now hypothesize that the giving a few-minute cooking range on most packages aims to accommodate the great sensitivity to the amount of added salt. On the other hand, our simple elastocapillarity experiment can accurately target a specific cooking time with a simple ruler measurement of the stick length between two hanging pasta noodles. Perhaps this experiment can further be validated by professional chefs and one day used by home cooks.

In summary, we have studied the capillary-induced coalescence of swelling pasta. We observed that, as pasta is cooked (hence swollen and softened), capillary action more easily deforms the noodle, and two pasta noodles can coalesce with the shorter stick length. Our predictive model that incorporates hygroscopic swelling and local elastic modulus as functions of the liquid concentration successfully captured the trend in the capillary coalescence of pasta. We believe that this work can provide an insight to manufacturers for determining an optimal cooking time for the most delectable pasta texture.

**Acknowledgements**
This work was supported by NSF CMMI (Grant No. 1825758).

**Competing financial interests**
Declared none

**Figures**

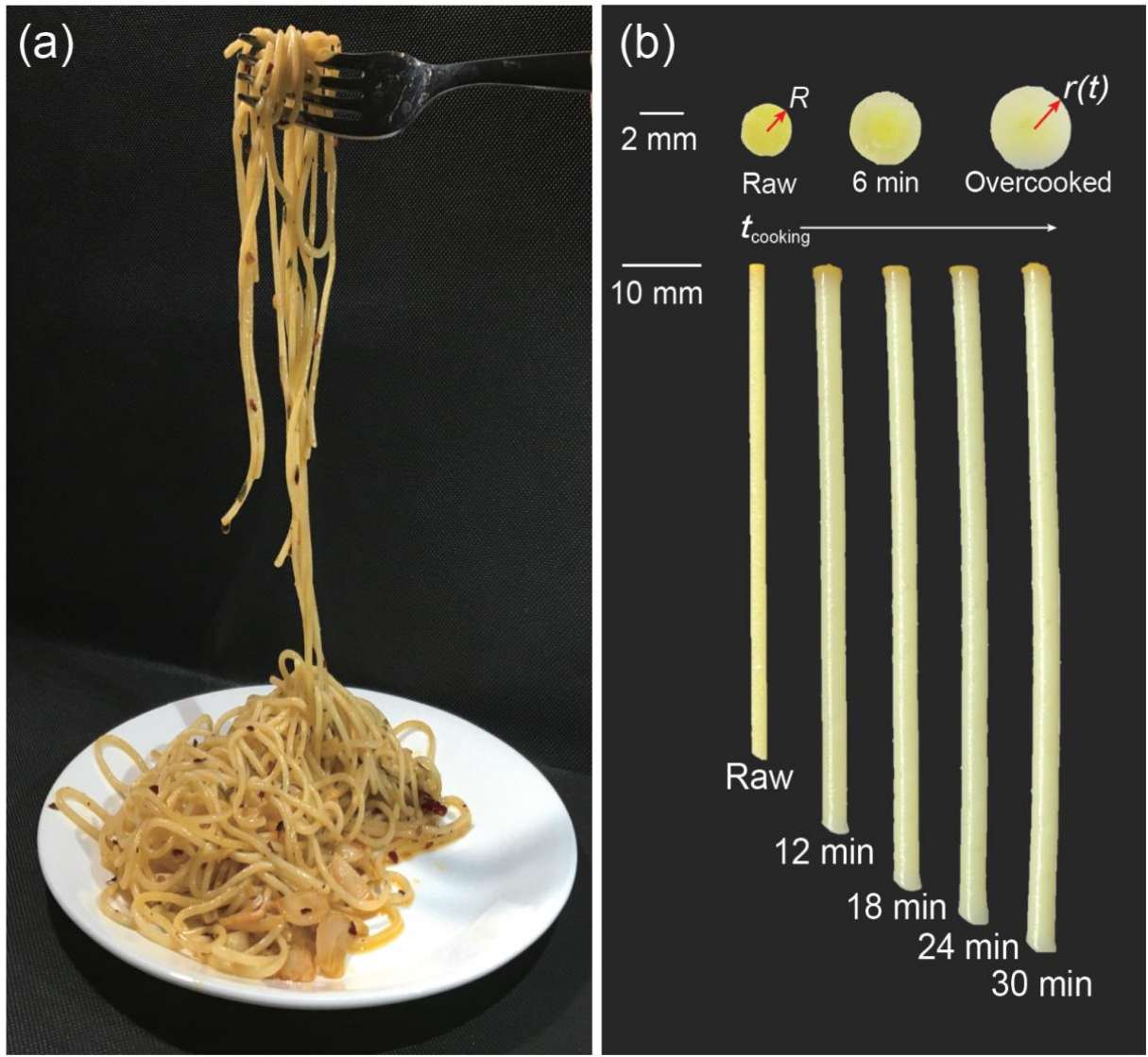

**Fig. 1**. (a) Picture of spaghetti aglio e olio. Capillary adhesion between the noodles is induced by olive oil. (b) Pasta radial and axial growth over time at 100 °C cooking due to hygroscopic swelling. Radial strain grows up to 70% of its original and axial strain up to 20% of its original after 30 minutes of cooking. Pasta cross-section shows color-gradient before it is sufficiently cooked.

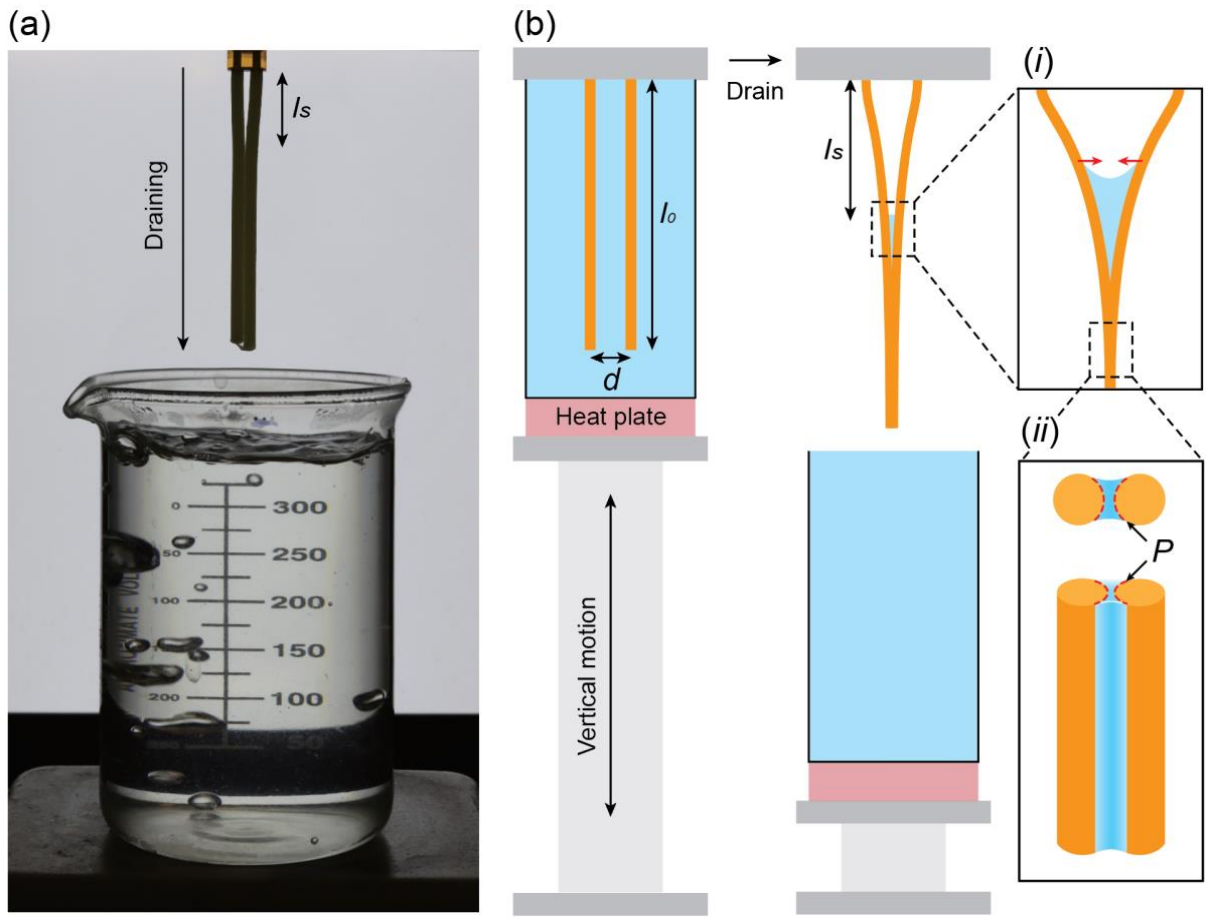

**Fig. 2**. Stick length measurement (a) proposed experimental setup for indirectly inferring the cooked pasta texture, for instance al dente, by measuring the stick length. Noodle stiction occurs as linear stage moves down the beaker. (b) Capillary adhesion mechanism analysis. Two pasta noodles displaced by a gap $d$ and cooked for every 3 minutes stick to each other when drained if capillary force is greater than the elastic restoration force. (i) liquid meniscus forms between the noodles that causes capillary adhesion (ii) pasta wetting situation seen from normal to the axial. We assume the liquid wets each half of the cylinder.

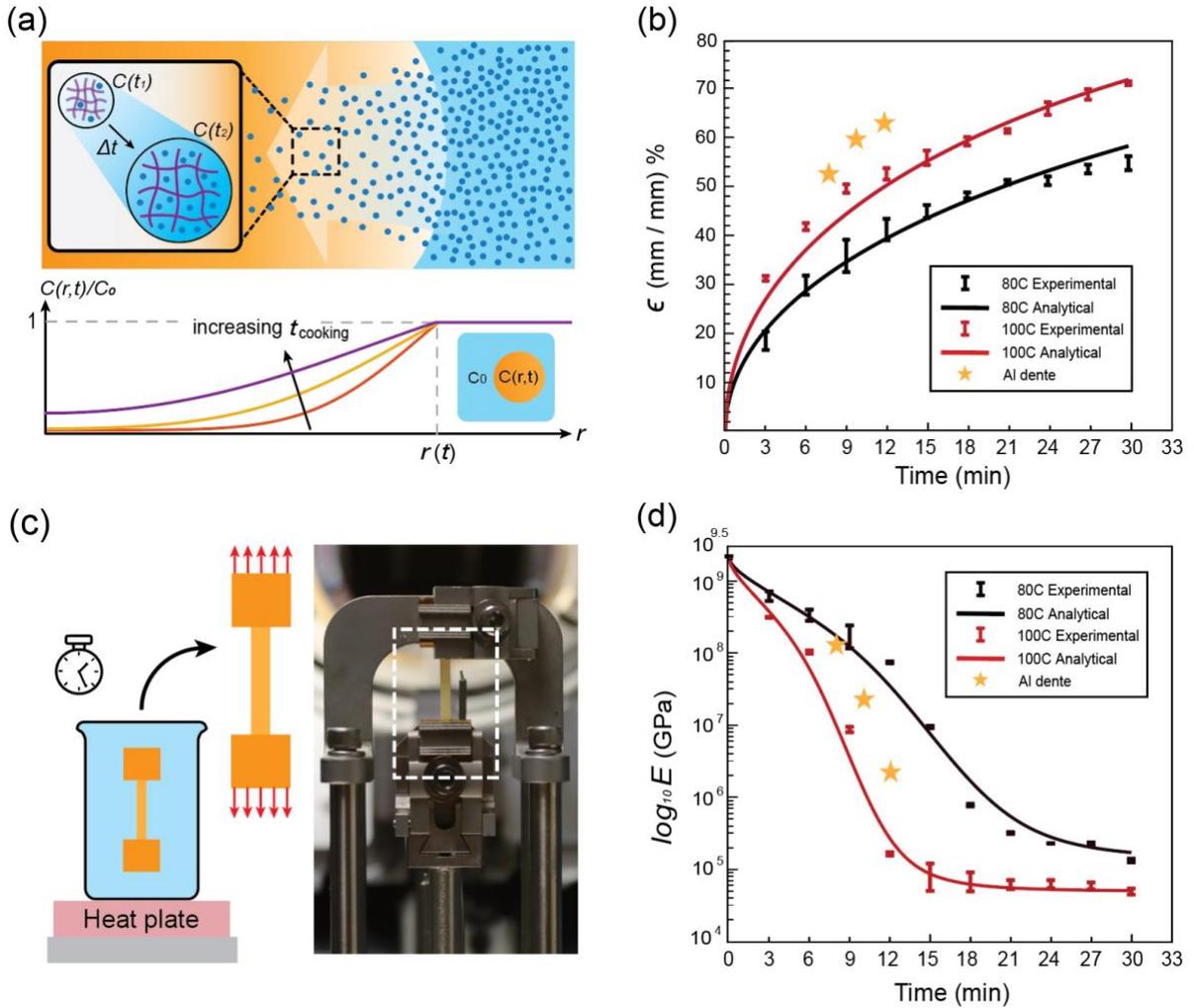

**Fig. 3**. (a) Qualitative description of hygroscopic swelling of pasta. Polymer matrix inside the pasta swells as liquid concentration within the material increases. The swelling is entirely due to the volume of liquid uptaken. $C(r,t)$ stands for local concentration and $C_0$ for concentration of surrounding liquid. Blue dots represent water molecules inside the macromolecular chain. (b) Experimental radial strain data and analytical solution for the radius over time. Coefficient of expansion is $\alpha = 0.9588$. Strain increases as cooking temperature or cooking time increases. Al dente samples (star symbols) reported higher strain compared to that of sample cooked at distilled water at the same temperature. (c) Young's modulus measurement setup. Sample modulus is measured with DMA after every 3 minutes of cooking. (d) Experimental Young's modulus data and analytical solution for the centroid of the modulus over time. Initial modulus is $E_0 = 2.17$ GPa, and the modulus drops by 5th order after sufficient cooking. The error bars of the plots are the standard deviation of the average strain of three measurements. We use 0.8 wt% salted water at 100 °C for Al dente (star symbols). Al dente samples (star symbols) reported higher modulus values compared to that of sample cooked at distilled water at the same temperature.

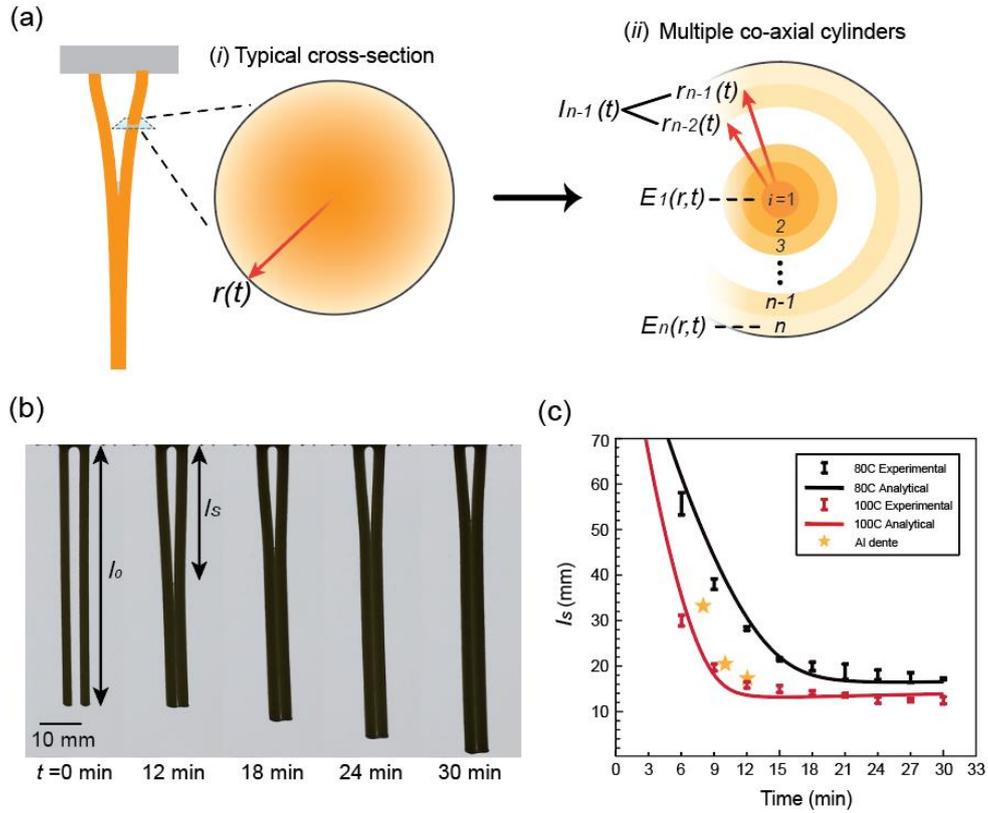

**Fig. 4**. (a) Qualitative description of the modeling of the flexural rigidity of the pasta. (i) Spatial gradient of the elastic modulus in the pasta cross section, normal to the vertical axis. Liquid concentration within the pasta increases gradually, so, modulus gradient exists radially. (ii) Mathematical model of modulus gradient. To model the rigidity, we divide the pasta into an infinite number of co-axial cylinders ($N \rightarrow \infty$). Using the local modulus and the timely growth of the pasta radius, one can obtain the rigidity of each hollow co-axial cylinders: $B_i = E_i I_i$. (b) Pictures of pasta stiction over cooking time. Stick length decreases over time while radius and length of the pasta increases (c) Experimental stick length data and analytical solution for the stick length over time. Stick length exponentially decreases until saturates. The error bars are the standard deviation of the average strain of # measurements. Stick length of Al dente samples increased compared to samples cooked at 100 °C without salt. It is mainly due to increased modulus. (see Fig. 3(d)).

# Supporting Information

# Swelling, Softening and Elastocapillary Adhesion of Cooked Pasta

Jonghyun Hwang[1*], Jonghyun Ha[1*], Ryan Siu[1], Yun Seong Kim[1,2], and Sameh Tawfick[1,2†]

[1]Department of Mechanical Science and Engineering, University of Illinois at Urbana-Champaign, Urbana, Illinois, 61801, USA
[2]The Beckman Institute for Advanced Science and Technology, University of Illinois at Urbana-Champaign, Urbana, Illinois, 61801, USA

*These authors contributed equally to the work
†Author to whom correspondence should be addressed. E-mail: tawfick@illinois.edu

**Note S1. Solution to diffusion equation and radial strain expression derivation**

Liquid migrates into the pasta, and it gradually swells and softens the polymer matrix. To express the local liquid concentration within the pasta, the following diffusion equation in cylindrical coordinates (Eqn. S1) should be solved:

$$\frac{\partial C}{\partial t} = D\left[\frac{1}{r}\frac{\partial}{\partial r}\left(r\frac{\partial C}{\partial r}\right) + \frac{1}{r^2}\frac{\partial^2 C}{\partial \theta^2} + \frac{\partial^2 C}{\partial z^2}\right]. \tag{S1}$$

Here, C is concentration or density of diffusing liquid as in scalar field and D is diffusivity. Assuming the pasta is a perfect cylinder, and the pasta length is much longer than its radius ($z \gg r$), the last two terms in RHS of S1 can be neglected. Hence, the following general solution for the liquid concentration can be written as a function of pasta radius and cooking time.

$$C(r,t) = \left[c_0 J_0\left(\sqrt{\lambda_n}\, r\right)\right]\left[c_1 e^{-D\lambda_n t}\right]. \tag{S2}$$

We have the following boundary and initial conditions. Assume pasta is initially fully dried: $C(r,t)|_{t=0} = 0$. $C(r,t)|_{r=R} = C_0$. Lastly, the concentration is symmetrical around the central axis: $\partial C/\partial r\,|_{r=0} = 0$.

Now, we have a full analytic solution for the concentration distribution:

$$\frac{C(r,t)}{C_0} = 1 - 2\sum_{n=1}^{\infty} \frac{J_0\left(x_n \frac{r}{R}\right)\cdot e^{-x_n^2 t/\tau}}{x_n \cdot J_1(x_n)}. \tag{S3}$$

Here, $\tau$ is a time constant expressed in terms of the radius, $R$, diffusivity, $D$, and a fitting parameter, $\beta$: $\tau = \beta R^2/D$. $J_0$ and $J_1$ are cylindrical Bessel functions of first kind of the zeroth and the first order, respectively. $x_n$ is the root of the zeroth cylindrical Bessel function of the first kind. Both the cylindrical Bessel functions are oscillating functions that decay over time. Because the system has *n* infinite eigenvalues, the expression is an infinite sum of the functions. We define the amount of liquid diffused into the cylinder with Eqn. S4.

$$V(t) = \int_0^{2\pi} \int_0^R C(r,t) r \, dr \, d\theta. \tag{S4}$$

At infinite $t$, the maximum amount of liquid diffusion equals: $V_\infty = \pi R^2 C_0$. The following expression is the nondimensional expression for the amount diffused into.

$$\frac{V(t)}{V_\infty} = 1 - 4 \sum_{n=1}^{\infty} \frac{e^{-x_n^2 t/\tau}}{x_n^2}. \tag{S5}$$

Using this relationship and the coefficient of expansion, $\alpha$, we can define the strain in the radial direction, as shown in Eqn. S6.

$$\epsilon_{rr}(t) = \alpha \left[\frac{V(t)}{V_\infty}\right]. \tag{S6}$$

Lastly, since the radius is expressed as $r(t) = R(1 + \epsilon)$, we get the following expression for the radius over time.

$$r(t) = R(1 + \epsilon_{rr}(t)) = R \left[1 + \alpha \left(1 - 4 \sum_{n=1}^{\infty} \frac{e^{-t(x_n^2/\tau)}}{x_n^2}\right)\right]. \tag{S7}$$

**Note S2. Local elastic modulus derivation using Weibull cumulative distribution**

Using the equation S3, which is the nondimensionalized local concentration, and by combining the expression to the Weibull cumulative distribution, we get the local Young's modulus, $E_i(r,t)$. (Eqn. S8). Since we are assuming a circular symmetry, $E_i$ is the modulus of the $i$ th co-axial cylinder.

$$\ln(E_i(r,t)) = \ln E_0 - [\ln E_0 - \ln E_s]\left(1 - e^{-\left(\frac{C(r,t)/C_0}{\eta}\right)^\theta}\right) \tag{S8}$$

Similar to Eqns S4 and S5, we can define the averaged modulus at that time.

$$E_{centroid}(t) = \frac{1}{\pi R^2} \int_0^{2\pi} \int_0^R E_i(r,t) r \, dr \, d\theta \tag{S9}$$

Since the experimentally measured Young's modulus is the averaged modulus of the composite, $E_{centroid}(t)$ should equal $E_{measured}(t)$. By fitting the centroid modulus to the experimental data, we back-solve for the two parameters of the Weibull modulus, the scale parameter, $\eta$, and the shape parameter, $\theta$. The scale parameter stretches/contracts the curve along the time axis, whereas the shape parameter affects the failure rate; $\theta$ being smaller than 1 represents that failure rate would degrease with time and vice versa.

Now that we have expressions for local elastic modulus and the radial strain, we can define the flexural rigidity of the pasta at any cooking time, $B(t)$, by assuming the cylinder is composed of infinite number of hollow cylinders that adds up to the whole, as shown in Eqn. S10.

$$B(t) = \sum_{i=0}^{\infty} E_i(r,t)[r_{i+1}(t)^4 - r_i(t)^4] \tag{S10}$$